\documentclass[10pt,twocolumn]{article}

\usepackage{graphicx}
\usepackage{epstopdf}
\usepackage{dcolumn}
\usepackage{bm}

\usepackage{amssymb}
\usepackage{amsmath}
\usepackage{amsfonts}

\newcommand{\chiqui}[1]{_{\mbox{\tiny{#1}}}}

\newcommand{\ti}{\'{\i}}

\newcommand{\parcial}[2]{\frac{\partial#1}{\partial#2}}

\newcommand{\metedos}{\varphi_{\Omega}}
\newcommand{\mb}{b_{\Omega}}
\newcommand{\ma}{a_{\Omega}}
\newcommand{\mkhh}{\chi_{\Omega}}

\newcommand{\under}{\underline}
\newcommand{\mitad}{\frac{1}{2}}
\newcommand{\e}{\epsilon}
\newcommand{\Om}{\Omega}
\newcommand{\om}{\omega}

\newcommand{\bn}{\under{n}}

\newcommand{\ket}[1]{\left|\, #1\, \right\rangle}
\newcommand{\espe}[3]{\left\langle\, #1\, \left|\, #2\, \right|\, #3\, \right\rangle}

\title{Euclidean Approach for Entropy of Black Shells}
\author{J. Robel Arenas S.\thanks{e-mail: jrarenass@unal.edu.co}\\ Observatorio Astron\'omico Nacional\\ Universidad Nacional de Colombia, Bogot\'a, Colombia \vspace{0.2cm} \\
Fernando Castro O.\thanks{e-mail: fcastroos@unal.edu.co}\\ Departamento de F\ti sica \\ Universidad Nacional de Colombia,  Bogot\'a, Colombia}

\begin{document}

\maketitle






\begin{abstract}
We introduce the concept of black shell, consisting on a  massive thin spherical shell contracting toward its gravitational radius from the point of view of an external observer far from the shell, in order to effectively model the gravitational collapse.  Considering  complementary description of entanglement entropy of a black shell and according to Gibbons-Hawking Euclidean approach, we calculate the Bekenstein-Hawking  entropy  retrieving  horizon integral and discarding boundary at infinity.\\

Key words: Black hole entropy, Entanglement entropy, Black hole thermodynamics.\\
PACS numbers: 04.70.Dy

\end{abstract}



\section{Introduction}
\label{intro}

Bekenstein-Hawking entropy  $S_{BH}$ has been derived from different points of view \cite{F}. But if it is considered thermal with a microscopic description,
according to the statistical foundations of  entropy, perhaps the most promising and appropriate formalism to explain $S_{BH}$ is the entanglement entropy approach \cite{ANC11,BS}. In particular, entanglement entropy of black shells is required, because the usual thermal entropies for black holes are divergent and geometric in nature \cite{F,BS}.

In the first part of this paper we complete the concept of black shell presented in \cite{BS}. Thus, by simplicity we effectively model the significant features of the gravitational collapse, in terms of a massive thin spherical collapsing shell with respect to an external observer.

In the second part of the text we consider Euclidean approach for entropy of black shells, thinking in its complementary description of entanglement entropy. Thermal entropy or entanglement entropy modeled by a black shell is a real physical model for $S_{BH}$, because it corresponds to the thermodynamics of hot quantum fields confined near the outside of the shell \cite{BW2,BS}. Unlike black holes that don't have structure and require a picture of quantum gravity to describe its geometrical entropy 
as some unknown physics, black shells have structure and thermal physical entropy.

We  show below that entropy calculated from Gibbons-Hawking Euclidean approach for spherical black shells, retrieving the horizon integral and discarding boundary at infinity, preserves
 the same expression  for Bekenstein-Hawking entropy:  
\begin{equation}\label{1}
S_{BH}=\frac{A}{4}\,.
\end{equation}

So entanglement entropy for black shells according to an external observer do not need to consider quantum gravity criteria to explain $S_{BH}$.

In the last part of the paper we complete the model by describing complementary quantum details.

 In Sec. 2 we present the black shell model and review the  Darmois-Israel formalism  \cite{DI,DI1} that is needed in order to obtain the motion equations of this model that we solve in Sec. 3. In Sec. 4 we reproduce the well known Bekenstein-Hawking entropy from Euclidean approach for a Schwarzschild black hole  \cite{EU1}. In Sec. 5 we retrieve the horizon integral and discard boundary at infinity in order to calculate the entropy of a black shell. There we introduce a mathematical identity to support our two complementary descriptions of physics near an event horizon. In Sec. 6, for completeness,  we return to entanglement entropy of black shells. So, we reproduce some results from  \cite{BS} and complete details of the corresponding  Hamiltonian formulation. Sec. 7 is devoted to discuss some conclusions  about this paper.

\section{Black shell model}

We present in this section a model where a black hole is formed by
a thin contracting shell of dust of mass $m$ that contracts beginning
from infinity and at last tends to the Schwarzschild radius from the
point of view of an external observer far from the shell in a gravitational
collapse evolution of Oppenheimer\textendash{}Snyder kind 
\cite{OS}, as we see in section 3.\\ In this context we employ the Darmois-Israel formalism  \cite{DI,DI1}
for a spherical thin shell $\Sigma$ which divides the spacetime in
two regions: the interior region $M^{-}$, described by flat Minkowskian
geometry and $M^{+}$, the exterior geometry described by Schwarzschild
spacetime. Both regions are described by spherical coordinates:
$X^{\alpha}=\left(t,r,\theta,\varphi\right)$, and we  use geometric
units in which: $C=G=1$, and  signature: $\left(-,+,+,+\right)$. Then the line element is:

\begin{equation}\label{BS1}
dS^{2}=-f\left(r\right)dt^{2}+f^{-1}\left(r\right)dr^{2}+r^{2}d\theta^{2}+r^{2}\sin^{2}\theta d\varphi^{2}.
\end{equation}

\medskip{}

Where: $f\left(r\right)=1-\frac{2m}{r}$ for $M^{+}$ and $f(r)=1$
for $M^{-}$.

\medskip{}

The shell hypersurface  $\sum$ is represented in the chosen frame by the following
parametric equation:

\medskip{}

\begin{equation}\label{F2}
h\left(r\right)=r-r_{0}=0.
\end{equation}

\medskip{}

where $r_{0}$ is the thin shell radius.

\medskip{}

The following intrinsic coordinates on the shell are used:\\  $\xi^{i}=\left(t,\theta,\varphi\right)$, and the 3-metric elements induced in the hypersurface are:

\medskip{}

\begin{equation}\label{F3}
g_{ij}=g_{\alpha\beta}\frac{\partial X^{\alpha}}{\partial\xi^{i}}\frac{\partial X^{\beta}}{\partial\xi^{j}}.
\end{equation}

\medskip{}

We note that the relation between coordinates of $M^{\pm}$ and
intrinsic coordinates on $\Sigma$ are:

\medskip{}

\begin{equation}\label{F4}
\frac{\partial X^{\alpha}}{\partial\xi^{i}}=\delta_{i}^{\alpha}.
\end{equation}

\medskip{}

and the parametric equation for $\sum$ adopts the following general form:

\medskip{}

\begin{equation}\label{F5}
h\left(x^{\alpha}\left(\xi^{i}\right)\right)=0.
\end{equation}

\medskip{}

We assume that $\sum$ is non-null and the unit 4-normals to $\sum$
in $M$ are:

\medskip{}

\begin{equation}\label{F6}
n_{\alpha}=\pm\frac{1}{\left(\mid g^{\beta\gamma}\frac{\partial h}{\partial x^{\beta}}\frac{\partial h}{\partial x^{\gamma}}\mid\right)^{\frac{1}{2}}}\frac{\partial h}{\partial x^{\alpha}}.
\end{equation}

\medskip{}

The extrinsic curvature (seccond fundamental form) is defined by:

\medskip{}

\begin{equation}\label{F7}
K_{ij}=\frac{\partial x^{\alpha}}{\partial\xi^{i}}\frac{\partial x^{\beta}}{\partial\xi^{j}}\nabla_{\alpha}n_{\beta}=-n_{\gamma}\left(\frac{\partial^{2}x^{\gamma}}{{\partial\xi^{i}}{\partial\xi^{j}}}+\Gamma_{\alpha\beta}^{\gamma}\frac{\partial x^{\alpha}}{\partial\xi^{i}}\frac{\partial x^{\beta}}{\partial\xi^{j}}\right).
\end{equation}

\medskip{}

Using (\ref{BS1}) and (\ref{F2}) the normal (\ref{F6}) is:

\medskip{}

\begin{equation}\label{F9}
n_{\alpha}=f^{-\frac{1}{2}}\delta_{\alpha}^{r}.
\end{equation}

\medskip{}

With the usual formula for the covariant derivative equations
(\ref{F4}) and (\ref{F7}) yields the simple relation:

\medskip{}

\begin{equation}\label{F10}
K_{ij}=-f^{-\frac{1}{2}}\Gamma_{ij}^{r}.
\end{equation}

\medskip{}

Raising indexes and considering that Schwarzschild metric is diagonal,
we arrive to the simple results:

\medskip{}

\begin{equation}\label{F11}
\left[K_{t}^{t}\right]=\frac{1}{2}f^{-\frac{1}{2}}\partial_{r}f,
\end{equation}




\begin{equation}\label{F12}
\left[K_{\theta}^{\theta}\right]=\left[K_{\varphi}^{\varphi}\right]=\frac{1}{r}f^{\frac{1}{2}}-\frac{1}{r},
\end{equation}

\medskip{}

\begin{equation}\label{F13}
\left[K\right]=f^{-\frac{1}{2}}(\frac{1}{2}\partial_{r}f+\frac{2}{r}f-f^{\frac{1}{2}}\frac{2}{r}).
\end{equation}

\medskip{}
Where square brackets denotes a discontinuity across the layer, i.e.,
$\left[f\right]=f^{+}-f^{-}$ . Our following task is to perform the calculations
to obtain the motion equations of the shell and the Bekenstein-Hawking entropy of the shell. This is developed in sections 3 and 5.

\section{Thin shell gravitational collapse}

In this section we review the thin shells junction formalism  \cite{DI,DI1}, 
in order to study the motion of a spherical shell of dust that contracts
beginning at rest from infinity as seen by a distant observer whose proper time is the Schwarzschild time $t$.

For this purpose we consider the surface energy tensor of a shell of dust:

\medskip{}

\begin{equation}\label{FC1}
S^{ab}=\sigma U^{a}U^{b}.
\end{equation}

\bigskip{}

With the condition: $U^{a}U_{a}=-1.$ And where $\sigma$ is the rest
mass surface density of the dust.

If we consider a trajectory $\xi^{a}(s)$ of an element of rest mass
on the shell, then the 3-vector associated $U^{a}=\frac{d\xi^{a}}{ds}$
is tangent to $\Sigma$.

For a spherical shell, the region $M^{+}$exterior to the shell have
the metric given by equation (2). And the region $M^{-}\\ $interior
to the shell is Minkowskian.

The equation of the shell is:

\medskip{}

\begin{equation}\label{FC2}
r=R(s).
\end{equation}

\bigskip{}

where $s$ is the proper time measured by the dust particles.

The equation of motion of the shell could be obtained from the Lanczos
equation   \cite{DI,DI1} :

\medskip{}

\begin{equation}\label{FC3}
S_{ij}=\frac{1}{8\pi}\left(\left[K_{ij}\right]-g_{ij}\left[K_{ab}g^{ab}\right]\right).
\end{equation}

\medskip{}

From equations (\ref{BS1}), (\ref{FC1}), (\ref{FC2}) and (\ref{FC3}) we obtain rearranging
therms:

\medskip{}

\begin{equation}\label{FC4} 
\frac{dR}{ds}=\pm\sqrt{\left(a+\frac{m}{2aR}\right)^{2}-1}.
\end{equation}

\bigskip{}

We are interested in the velocity of the shell as measured by an external
observer at rest at infinity, then we must replace:

\medskip{}

\begin{equation}\label{FC5}
ds=\sqrt{1-\frac{2m}{R}}dt.
\end{equation}

\medskip{}

If the shell is initially at rest:

\medskip{}

\begin{equation}\label{FC6}
\lim_{R\rightarrow\infty}\frac{dR}{dt}=0\Rightarrow a=1.
\end{equation}

\bigskip{}

Because gravity is attractive it produces a flux entering to the thin
Gaussian region that encloses the shell and by this reason we must
choose the negative sign. With these consederations we finally obtain:
\begin{equation}\label{FC7}
\frac{dR}{dt}=-\sqrt{\left(\frac{m}{R}\right)\left(1+\frac{m}{4R}\right)\left(1-\frac{2m}{R}\right)}.
\end{equation}

This equation is of difficult integration but fortunately we can do
the following good approximation:\\
\vspace{0.1cm} \\
$g\left(R\right)=\sqrt{\left(\frac{m}{R}\right)\left(1+\frac{m}{4R}\right)\left(1-\frac{2m}{R}\right)}$
\begin{equation}\label{FC8A}
\approx K\left(\frac{m}{R}\right)\left(1-\frac{2m}{R}\right).
\end{equation}

Where $K$ must be chosen in order to match these two expressions
for $g\left(R_{Max}\right)$, where $\frac{dg}{dR}\mid_{R_{Max}}=0.$
From this equation we obtain: $R_{Max}\approx3,886m$ and $K\approx3$.

Then the motion equation reduces to:
\begin{equation}\label{FC9}
\frac{dR}{dt}=-3\left(\frac{m}{R}\right)\left(1-\frac{2m}{R}\right).
\end{equation}
This equation could be integrated analytically to obtain:
\begin{equation}\label{FC10}
R\left(t\right)=1+\left(R_{0}-1\right)e^{\left(-\left[\frac{1}{2}\left(R^{2}\left(t\right)-R_{0}^{2}\right)+\left(R\left(t\right)-R_{0}\right)\right]\right)}e^{\left(-\frac{t}{\tau}\right)}.
\end{equation}

Where $\tau=\frac{4m}{3}$, $R_{0}$ is the initial radius and $R\left(t\right)$ and  $R_{0}$ are measured in units
of Schwarzschild radius.

Taking little variations of $R$ respect to $R_{0}$ we can obtain:

\medskip{}

\begin{equation}\label{FC11}
R\left(t\right)=1+\left(R_{0}-1\right)e^{\left(-\frac{t}{\tau}\right)}.
\end{equation}

\medskip{}

This important result sais that from the point of view of an external
observer at rest and far from the horizon, the shell approaches to
Schwarzschild radius asymptotically. This is that we call a black shell model
of a black hole.

For an observer comoving with the shell it is easy to see from  (\ref{FC5}), (\ref{FC7}), (\ref{FC8A})
and (\ref{FC9}) that:

\medskip{}

\begin{equation}\label{FC12}
\frac{dR}{ds}=-3\left(\frac{m}{R}\right).
\end{equation}

\bigskip{}

Integrating this equation we obtain:

\begin{equation}\label{FC13}
R\left(s\right)=\sqrt{R_{0}^{2}-6ms}.
\end{equation}

Then we see that the shell reaches the radius $R=0$  in the finite
time:

\begin{equation}\label{FC14}
s=\frac{R_{0}^{2}}{6m}.
\end{equation}
\\

This is that we call a black shell free gravitational collapse.

\section{Bekenstein-Hawking entropy from Euclidean approach}

In this section we review the statistical derivation of Bekenstein-Hawking entropy ($S_{BH}$) for stationary black holes, using analytic continuation to Euclidean sector and imposing a period on Euclidean time.

According to Gibbons-Hawking derivation \cite{EU1}, and in order to obtain the Bekenstein-Hawking entropy, we calculate the
action $I_{E}$ for the Schwarzschild metric (\ref{BS1}) :

\begin{equation}\label{FO1}
lnZ\approx I_{E}=\frac{1}{16\pi}\int_{M}R\sqrt{g}d^{4}x+\frac{1}{8\pi}\int_{\partial M}(K-K_{0}).
\end{equation}

Where $M$ is the Schwarzschild spacetime with (in natural units): $f\left(r\right)=1-\frac{2m}{r}$
, $R=0$ and $K$ is the second fundamental form.

Using the equations (\ref{F11}), (\ref{F12}) and (\ref{F13}) for $K_{0}=K\mid_{f=1}$ and
$r\rightarrow\infty$ we obtain:\\
 \vspace{0.1cm}\\ 
$lnZ=\frac{1}{8\pi}\int_{0}^{\beta}d\tau\int_{0}^{2\pi}d\varphi\int_{0}^{\pi}d\theta\left(f^{\frac{1}{2}}r^{2}\sin\theta\right)$
\begin{equation}\label{FO2A}
\times\left. \left(\frac{1}{2}\partial_{r}f+\frac{2}{r}f-\frac{2}{r}f^{\frac{1}{2}}\right) \right\vert_{r\rightarrow\infty}=-\frac{1}{2}\beta m,
\end{equation}

Where $\beta=8\pi m$ is the Euclidean period.

Then for the  black hole mass, $E=m$, we obtain:

\begin{equation}\label{FO3}
S_{BH}=lnZ+\beta E=4\pi m^{2}=\frac{1}{4}A.
\end{equation}

This is the Bekenstein-Hawking entropy of a black hole as was derived by Gibbons-Hawking \cite{EU1}.

\section{Zero-Loop Euclidean action for entropy of a Black Shell}

For a Black Shell, according to section 3 above,  the radius approaches to the Schwarzschild one and by this reason we 
retrieve the horizon integral obtaining the Euclidean action :\\
 \vspace{0.1cm} \\
$lnZ=\frac{1}{8\pi}\int_{0}^{\beta}d\tau\int_{0}^{2\pi}d\varphi\int_{0}^{\pi}d\theta\left(f^{\frac{1}{2}}r^{2}\sin\theta\right)$
\begin{equation}\label{EUC13DA}
\times\left. \left(\frac{1}{2}\partial_{r}f+\frac{2}{r}f-\frac{2}{r}f^{\frac{1}{2}}\right) \right\vert_{r\rightarrow2m}=\frac{1}{2}\beta m,
\end{equation}

where $\beta=8\pi m$ is the Euclidean period.

For a Black Shell the inner space is the empty space and by this
reason : $E=0$ obtaining:

\begin{equation}\label{EUC13E}
S_{BH}=lnZ+\beta E=4\pi m
^{2}=\frac{1}{4}A.
\end{equation}

Both results: (\ref{FO3}) and (\ref{EUC13E}), are mathematically equivalent derivations of $S_{BH}$. In order to relate both procedures, consider the following
mathematical identity:

\begin{equation}\label{EUC13F}
  \frac{1}{8\pi}\int_{\infty} (K - K_{\circ})\, d\Sigma + \frac{1}{8\pi} \int_{H} (K - K_{\circ})\, d\Sigma 
\end{equation}
\begin{equation}\label{EUC13FA}
 =\frac{1}{16\pi} \int_{r_{H} < r < \infty}R \sqrt{g} d^{4}x.
\end{equation}

This identity corresponds to a physical model that resolves one of the questions raised in Mukohyama-Israel \cite{BW2}, in the sense  that $S_{BH}$ is not a one-loop 
correction to the zero-loop Gibbons-Hawking contribution. Indeed this entropy is a zero-loop black shell contribution.

\section{Entanglement approach for entropy of black shells}

According to entanglement entropy model of black shells, thermal energy strongly concentrated near the exterior of a starlike object is clearly evident. This object has a reflecting surface, compressed to nearly (but not quite) its gravitational radius. Following this model we may approximate the total stress-energy (ground state and thermal excitations)
${(T_{\alpha \beta})}\chiqui{H}$, near the wall, to the Hartle-Hawking stress-energy ${(T_{\alpha \beta})}\chiqui{HH}$ \cite{BS,PVI}:

\begin{equation}
{(T_{\alpha \beta})}\chiqui{HH} \approx {(T_{\alpha \beta})}\chiqui{H} = {(T_{\alpha \beta})}\chiqui{B} + {(\Delta T_{\alpha \beta})}\chiqui{therm},
\end{equation}	

where ${(T_{\alpha \beta})}\chiqui{B}$ is Boulware stress tensor and ${(\Delta T_{\alpha \beta})}\chiqui{therm}$ are thermal excitations.

\subsection{Hamiltonian formulation}

\begin{figure}
\includegraphics{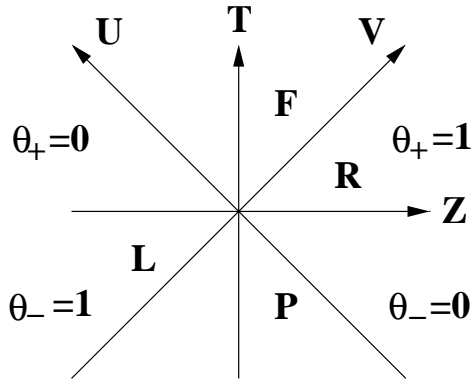}

\caption{Schwarzschild spacetime maximally extended}
\label{Mg10I}

\end{figure}

Thermofield dynamics encodes a reflexive symmetry between twin subsystems in such a way that each of them becomes macroscopically indistinguishable from a hot body at a definite temperature. The reflection symmetry of these subsystems corresponds to the right and left regions of an eternal black hole, i.e., Kruskal sectors R and L See (Fig.~\ref{Mg10I}). Thus, the twin subsystems correspond to field modes propagating in these (causally disjoint) kruskal sectors, and the thermally entangled state of the total system corresponds to the ground state on the full Kruskal manifold. In that sense, this section defines modes and ground states appropriate for the subsystems $L,R$ and for the global system (full Kruskal manifold), i.e., for stationary observers in a static spacetime (Killing-Boulware modes) and for free falling observers at the horizon of a black hole spacetime (Kruskal-Hartle-Hawking modes), and establishes the relationship between them.\\

From the line element (\ref{BS1}) consider the generic situation of a real scalar field $\Phi$ propagating on a geometrical background with static metric given by 

\begin{equation}\label{ETE1}
ds^{2} = g_{\chiqui{00}}\, dt^{2}+ g_{ab}\, dx^{a}\, dx^{b}\, .
\end{equation}

On the other hand, $\Phi$ may be expanded in terms of  Kruskal-Hartle-Hawking (KH$^{2}$)-modes $\mkhh^{(\epsilon)}(x)$ and Killing-Boulware (KB)-modes $\metedos^{(\epsilon)}(x)$, which are connected by the Bogoliubov transformation \cite{BS}

\begin{equation}\label{ETE30}
\mkhh^{(\epsilon)}(x) = \metedos^{(\epsilon)}(x)\, \cosh \chi + \metedos^
{(-\epsilon)}(x)\, \sinh \chi,
\end{equation}

with $\epsilon = \pm$.

The physical sense of (\ref{ETE30}) is based in the invariant action $S[\Phi]$ and invariant Hamiltonian $H$ under this transformation, according to the action

\begin{equation}\label{ETE33}
\mbox{\normalsize $S[\Phi] =\int \mathcal{L}[\Phi] d^{4}x =\int L[\Phi] dt_{+} =
\int_{-\infty}^{\infty}dt \left(\sum_{\epsilon}\epsilon
L^{(\epsilon)}(\Phi)\right),$} 
\end{equation}

\begin{figure*}

\includegraphics{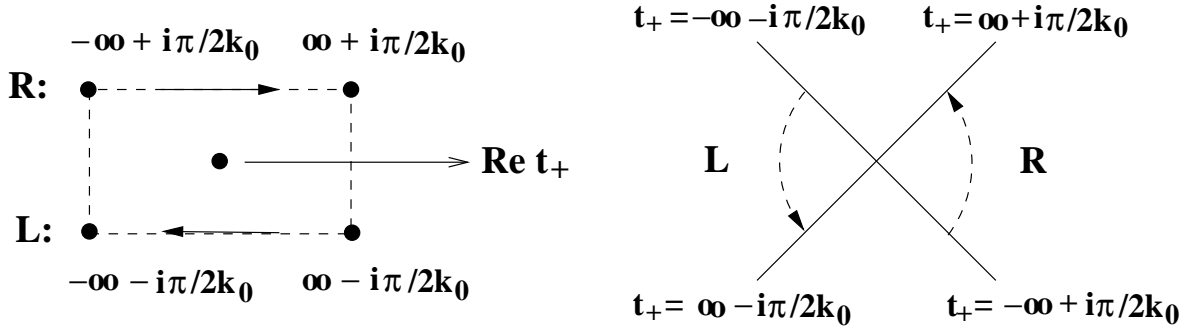}
\caption{Integration regions for the action}
\label{Mg10II}

\end{figure*} 

where the integration regions for the first and second integrals corresponds in Fig.~\ref{Mg10II} to the second and first graphs, respectively. There $\under{x}$  takes same value at two mirror points in  $L$ and $R$ sectors. These two points are distinguished by different imaginary parts of $t_{+}$:
\begin{equation}\label{ETE20}
 t_{+} =
\begin{cases}
t + \frac{i\pi}{2\kappa_{\chiqui{0}}}\, , \quad &x \in R \\
t - \frac{i\pi}{2\kappa_{\chiqui{0}}}\, , \quad &x \in L\, ,
\end{cases} 
\end{equation}
with $x \equiv (x^{\sigma}) = (\under{x}, t_{+})$ and  $\kappa_{\chiqui{0}}$, the surface gravity. $ t_{+}$ is analytic over the full Kruskal manifold, with the property that $e^{- i\alpha t_{+}}$ is positive frequency in Kruskal time.

$L$ sector contribution to $S$ enters with negative sign, because $t$ and $t_{+}$ run backwards in this sector.

Modes in the $L$ sector and beneath the horizon have negative energies. We are talking about the continuous extension of the Killing vector which is future timelike in the $R$ sector, and therefore becomes past-directed in the $L$ sector. Physically, this definition of energy includes gravitational potential energy, which is negative and becomes large below the horizon. In the Hartle-Hawking state, positive energies in the $R$ sector are evenly balanced with negative energies in the $L$ sector.

The state of the black hole is represented globally by the Hartle-Hawking state $\ket{0}_{\mbox{\tiny{H}}}$ \cite{BS}. The total energy of a quantum field in this state is zero, because of the balance between positive-energy modes in the R sector and negative-energy modes in the L sector. Formally, computing the expectation value of the global energy operator $\hat{H}$ in the state $\ket{0}_{\mbox{\tiny{H}}}$,
\begin{eqnarray}
\chiqui{H}\espe{0}{\hat{H}}{0}{\chiqui{H}}=&& Z^{-1}\sum_{\{n_{\om}\}}e^{-\beta E_{\bn}}\int_{0}^{\infty}d\om\, \om\, 
(-n_{\om} + n_{\om}) \notag \\
 \qquad =&&-\langle\, E\, \rangle_{\beta} + \langle\, E\, \rangle_{\beta} = 0\,
.\label{ETE88}
\end{eqnarray}

where
\begin{align}
\bn &\equiv \{n_{\Om}\, , \quad \text{for all  $\Om$ with $\om > 0$}\}, \notag \\
E_{\bn} &= \sum_{\substack{\Om \\ \om > 0}}n_{\Om}\, \om\, ,\quad
Z=\sum_{\bn}e^{-\beta E_{\bn}} = \prod_{\substack{\Om \\ \om > 0}}Z_{\Om}\,
. \label{ETE74}
\end{align}

On the other hand, the ground state $\ket{0}\chiqui{BR}$ for a parochial observer in $R$ is energetically depressed below the Hartle-Hawking ``vacuum''. If we form the expectation value of $\hat{H}$ in the state  $\ket{0}\chiqui{BR}$, only the L-sector part ($\e=-1$ in Eq.~(\ref{ETE54})) of $\hat{H}$ contributes, since there are no R-sector modes in this state. We thus find
\begin{equation}\label{ETE83}
\chiqui{BR}\espe{0}{\hat{H}}{0}{\chiqui{BR}} = - \langle\, E\, \rangle_{\beta}. 
\end{equation}
correspondingly, this same parochial observer, using his parochial energy operator $\hat{H}^{+}$, will perceive the Hartle-Hawking vacuum as energetically excited. If we form the expectation value of $\hat{H}^{+}$ in the state $\ket{0}_{\mbox{\tiny{H}}}$, the operator will pick out only the positive-energy modes in $R$, so
\begin{equation}\label{ETE87}
\chiqui{H}\espe{0}{\hat{H}^{+}}{0}{\chiqui{H}}
= +\langle\,  E\, \rangle_{\beta} ,
\end{equation}
where we have dropped zero-point energies.

\subsection{Entanglement entropy of black shells}

Entanglement entropy actually allows us to think that entropy arises physically located near the horizon, given by \cite{BS}

\begin{equation}
S = \int_{r_{0} + \e}^{R}4\pi\, r^{2}\, \frac{dr}{\sqrt{f}}\, s(r)\,
,\label{ETE191}
\end{equation}
where
\begin{equation}
s(r) = \frac{1}{3T^{2}}\int_{0}^{\infty} \frac{p^{2}\,
e^{\frac{E}{T}}}{(e^{\frac{E}{T}} - 1)^{2}}\, \frac{4\pi\, p^{2}\, dp}{h^{3}}\,
.\label{ETE192}
\end{equation}

The expressions above are similars to brick wall model \cite{BW1,BW2}, where $\e$ is defined by the external observer, which it is  the main difference with respect to brick wall model. In addition $s(r)$ is finite according to thermo field dynamics \cite{BS}.

According to this model, the integral~(\ref{ETE191}) is dominated by two contributions, for large $r=R$ and for small $r_{\chiqui{0}} + \e$. The former corresponds to a volume term, proportional to $\frac{4}{3}\pi r^{3}$, which represents the entropy and energy of a homogeneous quantum gas in a flat space at a uniform temperature $\frac{k_{0}}{2\pi}$. The latter is the contribution of gas near the inner wall $r=R_{\chiqui{0}}$. Then, for this last contribution is required to introduce the ultrarelativistic approximations
\begin{equation}\label{BW8}
s=\frac{4N}{\pi^{2}} T^{3} \text{,}\,\, \rho = \frac{3N}{\pi^{2}}T^{4}.
\end{equation}

Substituting Eq.~(\ref{BW8}) into Eq.~(\ref{ETE191}), the wall contribution to the total entropy is obtained
\begin{equation}\label{BW10}
S_{wall} = \frac{N}{90\pi \alpha^{2}} \frac{1}{4}A,
\end{equation}
where $N$ accounts for helicities and the number of particle species,  $A$ is the wall area and $\alpha$ is the proper radial distance from horizon to the shell.

Now, depending on $\alpha$, we can obtain the Bekenstein-Hawking entropy from Eq.~(\ref{BW10}) 
\begin{equation}\label{BW12}
S_{wall} = S\chiqui{BH}, 
\end{equation}
where $\alpha$ can be adjusted by resorting to times determined by an external observer in the context of the black shell model.\\

Unlike black holes, equation (\ref{BW10}) corresponds to a microscopic description in terms of quantum field modes.

\section{Discussions}

We have described in Sections 2 and 3 the motion of a spherical shell of dust that contracts beginning at rest from infinity by using  Darmois-Israel thin shells formalism \cite{DI,DI1}, then we obtain from the point of view of an external observer far from the horizon, that shell approaches to Schwarzschild radius. This result is one of the main characteristic of the black shell model, introduced above.\\

In sec. 4 we reproduce geometrical Bekenstein-Hawking entropy of a Schwarzschild black hole using the well known Gibbons-Hawking Euclidean approach  \cite{EU1} and derive in sec. 5 physical entropy of a black shell from Gibbons-Hawking Euclidean approach retrieving horizon integral and discarding boundary at infinity. This important result was obtained considering that for a black shell the mass is outside the horizon.\\

For the sake of completeness, in sec. 6 we return to entanglement entropy of black shells, because it is a complementary description of Gibbons-Hawking Euclidean approach for a black shell. So we have completed and extended the idea of black shell presented in \cite{BS}. In particular, we reproduce some results about it and complete details of the corresponding Hamiltonian formulation. \\

We agree with S. Mukohyama and W. Israel that entropy contributed by thermal excitations or entanglement is not a one loop correction to the zero-loop Gibbons-Hawking contribution. Actually we may consider these two entropy sources as equivalent but mutually exclusive descriptions of what is externally the same physical situation \cite{BW2}. In equation (\ref{EUC13F}) we observe that in the Euclidean sector of the black shell space-time there is an inner boundary, the black shell itself. Thus inner boundary contribution to the Euclidean action cancels that of the outer boundary at infinity. So the Gibbons-Hawking zero-loop contribution is zero in this sense.\\

With the right identification of the ground state, the back-reaction problem in 't Hooft brick wall model is resolved \cite{BW2}. In that sense for black shells, we show in (\ref{ETE88}) that the divergent parts cancel each other. The Boulware ground state contribution( which is energetically depressed below the vacuum) and thermal excitations cancel. \\

In summary we propose an effective model consisting on a massive thin spherical shell contracting toward its gravitational radius with respect to an external observer in order to describe 
significant features of a gravitational collapsing mass. The collapsing massive shell is compressed near its gravitational radius defining an natural cut-off between horizon and 
shell depending on external observer. From this model we can obtain a thermal and no divergent entanglement entropy that could explain $S_{BH}$. \\ \\ \\

\noindent
\textbf{Acknowledgements}\\ \\
We are indebted to Werner Israel, our collaborator of much of the work presented here, for many contributions, stimulating discussions and helpful comments on the manuscript.

\appendix

\section{Specific Lagrangian and Hamiltonian Formalism}

Specific Lagrangian $L$ is given by  
\begin{equation}
L = \sum_{\epsilon}\epsilon\, L^{(\epsilon)}(\Phi)\, ; \quad
L^{(\epsilon)}(\Phi) = \int \mathcal{L}^{(\epsilon)}(\Phi)\, d^{3}\under{x}\, ,
\label{ETE34}\\
\end{equation}
where
\begin{equation}\label{ETE35}
\mbox{\small $\mathcal{L}^{(\epsilon)}(\Phi) = \frac{\sqrt{-g}}{2} \left\{-g^{\chiqui{00}}
\Phi^{\chiqui{2}},_{t} - \left(g^{ab} \Phi,_{a} \Phi,_{b} +
m^{\chiqui{2}}\Phi^{\chiqui{2}}\right)\right\}\Theta_{\epsilon}(x).$}, 
\end{equation}

\begin{equation}\label{ETE17}
\Theta_{\epsilon}(x) \equiv \frac{1}{2}\, \left\{\Theta(-\epsilon U) + \Theta(\epsilon
V)\right\}\, . 
\end{equation}

$\Theta$ is the unit step function,  $t$ runs backwards in $L$ sector ($\epsilon = -1$) and $U,V$, are the Kruskal times.

The corresponding Hamiltonian $H$ is
\begin{equation}\label{ETE36}
H = \sum_{\epsilon}H^{(\epsilon)}(\Phi, \Pi)\, ,\quad H^{(\epsilon)} = \int
\mathcal{H}^{(\epsilon)}\, d^{3}\under{x}\, , 
\end{equation}
with
\begin{align}
\Pi(x) &= \parcial{\mathcal{L}}{\Phi,_{t}} = \gamma(\under{x})\, \epsilon(x)\,
 \Phi,_{t}\, ,\label{ETE337}\\
\gamma(\under{x}) &\equiv \sqrt{-g}\, \left(-g^{\chiqui{00}}\right)\, , \label{ETE23}\\
\mathcal{H}(\Phi) &= \Pi \, \Phi,_{t} - \mathcal{L}(\Phi) =
\mathcal{H}^{(+)}(\Phi) - \mathcal{H}^{(-)}(\Phi)\, ,\label{ETE38}\\
\mathcal{L}(\Phi) &= \mathcal{L}^{(+)}(\Phi) - \mathcal{L}^{(-)}(\Phi)\,,\label{ETE39}
\end{align}
\begin{eqnarray}
\mathcal{H}^{(\epsilon)}(\Phi, \Pi) = &&\frac{1}{2} \gamma^{\chiqui{-1}}
\Pi^{\chiqui{2}}\Theta_{\epsilon}(x) + \frac{1}{2} \sqrt{-g} \nonumber \\
&& \times \left(g^{ab} \Phi,_{a}\, \Phi,_{b} 
 + m^{\chiqui{2}}\Phi^{\chiqui{2}}\right)\Theta_{\epsilon}(x) .\label{ETE40} 
\end{eqnarray}

An important result is the following:
both vacuum states, $\ket{0}_{\mbox{\tiny{H}}}$ and $\ket{0}_{\mbox{\tiny{B}}}$, have zero energy,
\begin{equation}\label{ETE75}
H\ket{0}_{\mbox{\tiny{H}}} = H\ket{0}_{\mbox{\tiny{B}}} = 0\, ,
\end{equation}
with $H$  given by the expressions 

\begin{equation}\label{ETE54}
H[\Phi, \Pi] = \sum_{\Om, \e}\e\, \mitad\, |\om|\left(\mb^{(\e)\dag}\,
\mb^{(\e)} + \mb^{(\e)}\, \mb^{(\e)\dag}\right)\, ,
\end{equation}

\begin{equation}\label{ETE57}
\sum_{\e} \e\, \ma^{(\e)\dag}\, \ma^{(\e)} = \sum_{\e}\e\, \mb^{(\e)\dag}\,
\mb^{(\e)}\, ,
\end{equation}

\begin{equation}\label{ETE58}
H[\Phi, \Pi] = \sum_{\Om, \e}\e\, \mitad\, |\om|\left(\ma^{(\e)\dag}\,
\ma^{(\e)} + \ma^{(\e)}\, \ma^{(\e)\dag}\right).
\end{equation}

\end{document}